\documentstyle[12pt,epsf]{article}  
\newcommand{\VV}{{\cal V}}
\newcommand{\VVV}{\wt{\cal V}}

\newcommand{\TT}{{\cal T}}

\newcommand{\wt}{\widetilde}
\newcommand{\wh}{\widehat}

\newcommand{\wb}{\bar}

\newcommand{\aon}{a}
\newcommand{\atw}{{\wb a}}
\newcommand{\bon}{b}
\newcommand{\btw}{{\wb b}}
\newcommand{\bth}{\wh b}
\newcommand{\fon}{f}
\newcommand{\ftw}{{\wb f}}
\newcommand{\ron}{r}
\newcommand{\rtw}{{\wb r}}

\newcommand{\be}{\begin{equation}}
\newcommand{\ee}{\end{equation}}
\newcommand{\ben}{\begin{eqnarray}\displaystyle}
\newcommand{\een}{\end{eqnarray}}
\newcommand{\refb}[1]{(\ref{#1})}

\newcommand{\sectiono}[1]{\section{#1}\setcounter{equation}{0}}

\begin{document}

{}~ \hfill\vbox{\hbox{hep-th/0007153}\hbox{MRI-P-000703}
\hbox{CTP-MIT-3002}}\break

\vskip 3.0cm

\centerline{\large \bf Large Marginal Deformations in String Field Theory}

\vspace*{6.0ex}
\centerline{\large \rm Ashoke Sen
\footnote{E-mail: asen@thwgs.cern.ch, sen@mri.ernet.in}}

\vspace*{1.5ex}

\centerline{\large \it Mehta Research Institute of Mathematics}
 \centerline{\large \it and Mathematical Physics}

\centerline{\large \it  Chhatnag Road, Jhoosi,
Allahabad 211019, INDIA}

\vspace*{1.5ex}

\centerline{and}

\vspace{1.5ex}

\centerline{\large \rm Barton Zwiebach
\footnote{E-mail: zwiebach@mitlns.mit.edu}}

\vspace*{1.5ex}

\centerline{\large \it Center for Theoretical Physics}
\centerline{\large \it Massachussetts Institute of Technology}
\centerline{\large \it  Cambridge, MA 02139, USA}

\vspace*{3.5ex}
\medskip
\centerline {\bf Abstract}

\bigskip

We use the level truncation scheme to obtain accurate 
descriptions of open bosonic string field configurations 
 corresponding to large marginal deformations 
such as background Wilson lines.
To do so, we solve for all fields as functions of the
massless string field, and confirm that the
effective potential of the massless field becomes increasingly flat as
the level of approximation is increased. Surprisingly, as a result of
the merging of two branches of the solution - one originating at zero
tachyon vev and the other originating at the tachyonic vacuum -
this effective potential exists only for
a finite range of values of the massless field.
We use the D1 to D0 brane marginal transition on a circle to explore
the possibility  that this finite range corresponds
to the infinite range of the conformal field theory parameter
describing marginal deformations, but are unable to arrive
at a definitive conclusion.

\vfill \eject

\baselineskip=18pt

\tableofcontents

\sectiono{Introduction and Summary} \label{s1}

It has been realised recently that string field
theory (SFT)~\cite{WITTENBSFT,9503099,9912120} 
provides a useful
tool for
studying the phenomenon of tachyon condensation in string
theory~\cite{9912249,0001201,0002237,0001084,0002211,0003220,0004015} using
the level truncation scheme developed in ref.\cite{KS}. This includes the
open string tachyon on a D-brane of bosonic string theory, as well as on a
non-BPS D-brane or a D-brane anti-D-brane pair in type II string theory.
The analysis has been extended to discuss condensation of modes of the
tachyon carrying momentum along compact directions as long as the
effective (mass)$^2$ of this mode remains 
negative~\cite{0002117,0003031,0005036}. In the language of two dimensional
conformal field theory (CFT) describing the propagation of the string,
switching
on a vacuum expectation value (vev) for such tachyonic fields corresponds to
switching on relevant perturbations. Although conformal field theory
analysis has been used to gain insight into the phenomenon of tachyon
condensation in some of these cases~\cite{0003101,9406125}, string field
theory certainly seems to provide a unified approach to the study of these
phenomena. 
An alternative approach to the study of these phenomena based
on effective field theory on non-commutative spaces has been proposed
in refs.\cite{0003160,0005006,0005031,0007078}, 
and its
relationship to
string
field
theory has been
discussed in \cite{0006071}.

Typically D-branes also contain massless open string states. In many cases
the potential for the fields associated with these modes has exact flat
directions. Switching on a vev for these fields corresponds to exactly
marginal deformations of the corresponding conformal field theory. A
typical example of such a marginal deformation is the Wilson line;
a constant vev of a U(1) gauge field along a compact
direction. These deformations can be studied using well known techniques
of conformal field theory. On the other hand, studying these in string
field theory seems to be a difficult problem due to the following reason.
Whereas we expect that the exact potential in string field theory will
have an exact flat direction corresponding to each marginal deformation of
the two dimensional conformal field theory, to any given order in the
level truncation scheme the potential is not exactly flat. As a result,
if we try to solve the equations of motion of string field theory using
the level truncation scheme, then instead of getting a one parameter
family of solutions corresponding to each marginal direction, we get
isolated solutions.

This is the problem that we address in this paper. We take our marginal
deformation parameter $a_{s}$, where the $s$ stands for SFT, 
to be that associated with the constant vev of a U(1) 
gauge field. Instead of trying to solve the equations of motion of all
components of the string field, we hold fixed 
$a_s$ 
and solve for all other fields as a function of $a_s$ 
using their equations of motion. This allows us to find the `effective
potential' for the massless field $a_s$. 
At any given order in the level
expansion, this potential is not flat, and hence it has at most isolated
extrema. 
We find, however, that as we increase the level of approximation,
the potential becomes flatter, 
strongly suggesting 
that the exact effective
potential is indeed flat.  
This shows that for
a fixed vev of
the  massless field $a_s$, the
solution of the equations of motion of all other fields gives an accurate
representation of the string field configuration corresponding to the
deformed conformal field theory.
The flatness of the effective potential
is consistent with the earlier result of Taylor \cite{0001201}
showing  that the coefficient of the
leading quartic term in the expansion of the
effective potential around the origin
does approach zero
as the level of approximation is increased.

During the process of determining the effective potential of the marginal
field, we find a surprise: the effective potential exists only if 
$|a_s|$  
is less than a certain value $\overline a_s$. 
This restriction can be
understood as follows. For zero vev of the marginal field, the equations
of motion of the other fields have at least two solutions: the trivial
solution where all other fields vanish, and the tachyonic vacuum solution
studied in \cite{KS,9912249,0002237}. We shall refer to these two
solutions as the $M$-branch, 
for marginal, and  the $V$-branch for vacuum, respectively. 
When we switch on a small vev $a_s$ of the marginal 
field, the 
two branches 
still remain,
although the vev of the other fields associated with 
these two branches change slightly.
For the study of the tachyonic vacuum, the $V$-branch is the relevant 
branch, but for
our study, the $M$-branch is the relevant branch. We find that as we 
increase\footnote{Due to  
the existence of a $Z_2$ symmetry under which $a_s\to -a_s$, we can
restrict our study to positive values of $a_s$ only.}
the value of $a_s$
these two branches come closer
together, and at certain critical value of the vev, they meet. Beyond this
point there are no real solutions associated to these branches. This
phenomenon can be seen analytically at the level (1,2) approximation to
the potential, but we have checked numerically that this phenomenon
persists at least up to the level (4,8) approximation. Furthermore, the
vev of the massless field $a_s$ at which the two branches meet seems to
converge rapidly to a finite value $\overline a_s$ as we increase the level
of
approximation. We shall refer to 
$\overline a_s$  as the critical value of the string
field marginal parameter. 
This value, together with the values taken by the other
fields in the theory define the {\it critical} string field.

This brings us to the question of interpretation 
of the critical value $\overline a_s$.  
Although this corresponds to a finite string field configuration, there is
{\it a priori} no guarantee that it corresponds to a finite vev $a_c$
($c$ for CFT parameter) 
of the gauge field which appears {\it e.g.} in the Born-Infeld action. 
There are now two possibilities: 
\medskip

(1) The critical value $\overline a_s$ corresponds to
$a_c = \infty$, or

(2) The critical value $\overline a_s$ corresponds to a
finite value $\overline a_c$ of the gauge field vev. 

\medskip
\noindent
Unfortunately, with the
information available at present, we are unable to distinguish between
these two possibilities. If the first possibility holds, then this implies
that an infinite distance in the CFT moduli space (as measured in
Zamolodchikov metric) correponds to a finite shift in the string field. 
On the other hand, if the second possibility holds, then this would mean
that
open string field theory in a single coordinate system is unable to
describe the full CFT moduli space. 
However, we can describe the full CFT
moduli space by taking open string field 
theory in different coordinate patches. 

To see that this can be done,
note that all states are neutral
under the gauge field and therefore the
presence of the Wilson line does not affect the form of the
correlation functions of  the conformal
field theory for a suitable choice of basis of states in the Hilbert
space.
Thus, the string field theory action has exactly 
the same form about any background Wilson line,
and hence can span a given (finite) range of values of 
the Wilson line centered around the background value.
Thus clearly the whole range of Wilson line vev can
be spanned by putting together a set of string 
field theories formulated around
different background values of the Wilson line. 
The fields in two such
string field theories formulated around different 
background Wilson lines
are presumably related by complicated non-linear 
field redefinitions. In
the case of infinitesimal marginal deformations, 
these field redefinitions
were worked out in ref.\cite{9307088}.

We are not only unable to decide between options (1) and (2)
but it is also not clear to us whether or not the same phenomenon,
{\it i.e.} the appearance of a critical value $\overline a_s$,
also occurs in superstring field theory describing a
single BPS D-brane configuration. 
In this case there 
is no analog of a non-trivial tachyonic vacuum from which another branch
of the solution can originate,
and therefore a critical value would have to arise by a 
different mechanism, possibly involving the massive 
string fields. 
Although there are tachyonic modes on a non-BPS D-brane or a brane-antibrane
pair, the 
vev of the gauge field does not induce a tachyon vev due 
to the  GSO
$Z_2$ symmetry under which the tachyon is odd and the gauge field
is even. 
In fact, no state in the GSO odd sector will acquire 
a vev. 
Hence the above comments for the 
BPS brane apply to these cases as well.

Although we carry out the analysis in the specific case of marginal
deformations associated with the vev of a Wilson line, our results can be
used in a more general context. First of all, instead of considering a gauge
field vev along a direction tangential to the brane, we could consider
giving a vev to the scalar field representing translation of a brane along a
direction transverse to the brane. The effective potential for this mode
will be identical to that of the Wilson line, and hence all our results
apply to this mode. More generally, if we consider a
situation where the bulk 
$c=26$ matter conformal field theory has a U(1) 
current algebra, with the U(1) current satisfying either Dirichlet or
Neumann boundary condition at the boundary of the world-sheet, then the
U(1) current at the boundary of the world-sheet
corresponds to a marginal operator, and our result can be used to describe
the string field configuration corresponding to switching on this marginal
deformation. In particular, this includes deformations which create a
tachyonic lump on a circle of unit radius. As discussed in
refs.\cite{RECK,9902105,0003101,9406125}, if we consider a 
D-$p$-brane of bosonic string
theory with one of its tangential directions $x$ compactified on a circle
of radius $R$, and switch on background tachyon field proportional to
$\cos(x/R)$, the conformal field theory flows to that describing a
D-$(p-1)$-brane for $R\ge 1$. 
For $R=1$ this describes a marginal deformation, since
the tachyon vertex operator can be mapped to the boundary value of a U(1)
current\cite{9902105}. Hence the effective potential for this mode 
of the tachyon can be determined from our general
results. We verify this explicitly by determining the tachyon potential to
level (4,8) for arbitrary $R$ close to but larger than one, and then showing
that in the $R\to 1$ limit, this potential reduces to the 
level (4,8) potential
involving the Wilson line. Using the results for the tachyon potential at
level (4,8) approximation we also obtain a more accurate description of
the lump solution for $R$ close to but larger than one. This generalizes and
extends the analysis of ref.\cite{0005036}.  

The earlier analysis of 
\cite{9902105} shows that for $R$ close to but slightly larger than one, the
effective tachyon potential has a 
minimum at $a_c=\pm {1\over 2 \sqrt 2}$, 
corresponding to
turning the D-$p$ brane
into
a D-$(p-1)$ brane.\footnote{$a_c$ is normalized such that it multiplies 
a vertex operator of unit norm in the CFT action.} 
We use our present string field theory 
analysis of the tachyon potential
for $R$ near one to attempt to find the expectation value of the
string field theory variable $a_s$ 
representing the same minimum. This gives the values of $a_s$
corresponding
to $a_c=\pm {1\over 2\sqrt 2}$. 
We use this to try to gain more insight
into
the functional relationship between $a_c$ and $a_s$.
Unfortunately this analysis does not quite answer the question as to
whether $\overline a_s$ corresponds to finite or infinite value of $a_c$.

The rest of the paper is organised as follows. In section \ref{s2} we
construct the relevant part of the string field theory action needed for
computing the effective potential for the Wilson line, and use this to
compute
the effective potential. The computation is done analytically at level
(1,2) approximation and numerically up to level (4,8) approximation. In
section \ref{s3} we construct the string field theory action relevant for
studying the lump solution on a circle of radius $R$, and discuss its
equivalence with
the action of section \ref{s2} for $R=1$. We also use this potential for
$R>1$ to estimate the value of $a_s$ for $a_c={1\over 2\sqrt{2}}$. This is
used in section \ref{s4} to discuss the possible functional relation
between $a_s$ and $a_c$, in particular whether the upper limit on $a_s$
corresponds to finite or infinite value of $a_c$. We conclude in section
\ref{s5} with some comments. Appendices \ref{a1} and \ref{a2} contains the
details of the string field theory action relevant for the analysis of
sections \ref{s2} and \ref{s3} respectively.

\sectiono{String Fields for Wilson Line Marginal Deformations} \label{s2} 

In this section
we begin our analysis of string fields corresponding
to CFT marginal deformations. 
The setup is that of bosonic
open string field theory describing the 
dynamics of a D-$p$ brane. We single out
a particular coordinate $x$ along the world volume
of the brane and consider giving expectation value
to the constant mode of the gauge field component
$A_x$. This represents a marginal deformation of the BCFT
describing the D-brane. Our aim is to find the 
string field corresponding to such deformations. We will
not assume that 
the deformation is small. For our 
present analysis it will make no difference
whether or not the $x$-direction is compact; we give
$x$-independent expectation values to all fields, so that 
modes carrying non-zero momentum along $x$ 
(or any other direction) are set to zero.

We begin   
by examining the lowest level approximation to the problem where
happily, many of the features of the problem are already 
apparent. Then we
discuss its generalization to higher levels. 

\subsection{Lowest level analysis: Level (1,2)}

At the lowest level (level (1,2)) approximation we must include the
tachyon and the gauge field. The string field is therefore
\be
\label{sf}
|\Phi^{(1)}\rangle = ( t_0  + a_s \, \alpha_{-1}^X)\, c_1 |0\rangle \,, 
\ee
where $t_0$ denotes the level zero tachyon zero mode and
$a_s$ denotes the level one gauge field zero mode. $\alpha^X_n$ denotes
the $n$th oscillator mode of $X$. 
We now evaluate the string field action to get the potential
${\cal V}(\Phi) ( \equiv - S(\Phi) / (2 \pi^2 \TT_p)$ where $\TT_p$ is
the tension of the D-brane) associated to this string field. 
A small calculation gives\footnote{One can use, for example, the
conservation method of ref.\cite{0006240}
to find the $t_0 a_s^2$ coupling. Using equation (4.18) one finds 
that $\langle c_1, \alpha_{-1}^X c_1, 
\alpha_{-1}^X c_1 \rangle  = {16\over 27} \langle
c_1, c_1, c_1\rangle = {16\over 27} K^3 = K$.} 
\be
\label{fgh}
\VV (t_0, a_s) = -{1\over 2} t_0^2 
+ {1\over 3} K^3 t_0^3 + K t_0 a_s^2 \,,
\ee
where $K = 3\sqrt{3}/4$. We note the absence of a 
quadratic and cubic term for $a_s$. This, of course, is
expected by the standard CFT constraints on marginal
operators. Note that the sign of the last term is the
same as that of the second term. This will play a role
in what follows, and is a result of the positive norm
of the state $\alpha_{-1}^X |0\rangle$. 

We now find the effective potential for $a_s$ by integrating
out classically the tachyon field $t_0$. Since the equation
of motion is quadratic we find two solutions
\be
t_0^M =  {4\over 81\sqrt{3}} \Bigl( -\sqrt{64- 
729 a_s^2\,} + 8 \Bigr), \qquad
t_0^V =  {4\over 81\sqrt{3}} \Bigl( \sqrt{64- 729 a_s^2\,} + 8 \Bigr)\, .   
\ee
These are the marginal $M$-branch solution and the vacuum $V$-branch
solution respectively. Indeed one can see that for $a_s=0$ 
we get $t_0^M =0$, which is the 
value for the tachyon at the maximum of 
the potential, while
$t_0^V$ is the familiar tachyon expectation value at the 
local minimum 
of the cubic potential. It is clear 
from the above equations that
there are no real solutions for $t_0$ unless
\be
\label{cra}
|a_s| \leq {8\over 27} \equiv \overline a_s^{(1,2)}\,.  
\ee
Note that at this point the two branches for $t_0$ meet.

It is of interest to understand the nature of the effective 
potential for $a_s$.  Substituting the values of $t_0$ into
$\VV(t_0, a_s)$  and letting $\VV^{M/V}(a_s ) \equiv \VV(t_0^{M/V} (a_s),
a_s)$
we obtain:
\be
\label{epota}
\VV^{M/V} (a_s) = {2\over 59049} \Bigl( - 512 + 8748 a_s^2 
\pm (64 - 729 a_s^2)^{3/2}\Bigr)\,,
\ee
where the top sign (the $+$) goes for the $M$-branch, and the
bottom sign is for the $V$-branch.  We are particularly interested
in the $M$-branch, where we expect to have $a_s$ represent
a marginal direction. Thus ideally $\VV^M(a_s)$ should have been
identically zero. While it is not zero, the function is certainly 
relatively flat. It is a monotonically increasing function, 
and $2\pi^2 \TT_p \VV^M(\overline a_s^{(1,2)}) = 0.17\, \TT_p$, indicating
that
at
the end of the domain of definition the potential energy for the 
marginal direction fails to be zero by about 17\% of the D-brane
tension. We can
expand $\VV^M(a_s)$ for small
$a_s$ finding: 
\be 
\label{sas}
\VV^M(a_s) = {27\over 32} a_s^4 + {6561\over 4096} a_s^6 + \cdots \,. 
\ee
We shall see that as the level of 
approximation is increased the
potential $\VV^M$ becomes flatter. The leading coefficient 
(as well as the other expansion coefficients) will 
become smaller. 
The computation of the leading quartic term in the above
equation
is equivalent to the computation of the quartic gauge field interaction in
ref.~\cite{0001201}.

We can also examine the $V$-branch for $a_s$ small. We get
\be
\label{sbas} 
\VV^V(a_s) = -{2048\over 59049} + {16\over 27} a_s^2  + \cdots \,. 
\ee
Notice the leading constant term. 
The numerical value of $2 \pi^2 \TT_p \VV^V(a_s=0)$ is
$-.68 \TT_p$.
This is 68\% of the energy density of the original brane, and approaches
100\% of the brane tension as we increase the level of
approximation\cite{9912249,0002237}. Note
also that the quadratic
term for $a_s$ does not vanish (and will not become smaller as the level
of
approximation is increased).  This is consistent with the
expectation that there are no massless states around the stable
vacuum. The marginal direction has been lifted.\footnote{We thank
W. Taylor for raising the question of the fate of this marginal
direction on the stable vacuum, and for comparing with us
his results to be published independently \cite{taylor}.}

\begin{figure}[!ht]
\leavevmode
\begin{center}
\epsfbox{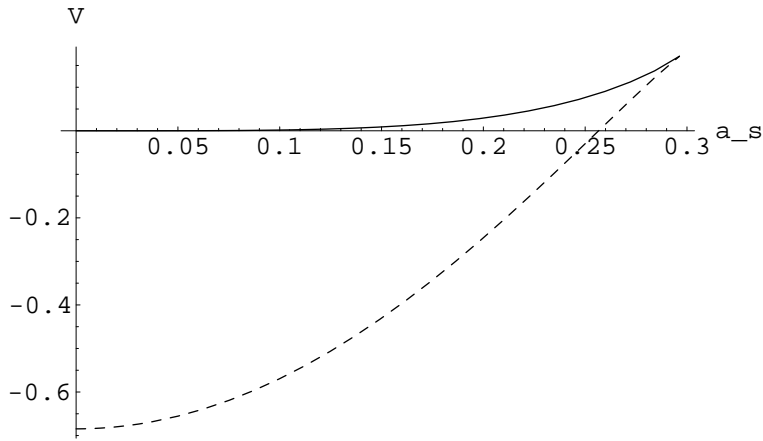}
\end{center}
\caption[]{\small The level (1,2) effective potential 
$V\equiv (2\pi^2 \VV^{M/V}(a_s))$ for 
$a_s$ on
the $M$-branch (solid line) and 
$V$-branch (dashed line). The two branches meet
at $a_s^{(1,2)} = 8/27$.} \label{f1}
\end{figure}

\subsection{Higher level analysis: Up to Level (4,8)}

We gave before the fields necessary to compute the
effective potential of the marginal parameter to level (1,2).
This included the tachyon and the level one field
$\alpha_{-1}^X c_1|0\rangle$ (see \refb{sft1}). Note that
the latter state 
is odd under both the twist symmetry (the twist 
eigenvalue is $\Omega = (-)^l$, with $l$ the level), and the
$X\to - X$ transformation. Thus it is even under the combined operation of
the 
twist and parity transformation. Since this is a symmetry of the string
field theory action~\cite{9705038}, it is clear that we can get a
consistent solution of
the full string field theory equations of motion by setting to zero fields
which are odd under this combined operation. Thus in extending our
analysis to higher level, we only need to include string field
configurations which are even under the combined operation of the twist
and the $X\to -X$ transformation. 
We shall further restrict the string field
configuration by including only the zero momentum modes,
using the Siegel gauge condition, and considerations involving closure
of the $*$-product algebra 
of a subset of a string fields along the lines 
discussed in \cite{9911116,9912249,0005036}. This amounts to including
states of ghost number one, obtained by acting on $c_1|0\rangle$ with
$b_{-n}$, $c_{-n}$, $\alpha^X_{-n}$ and $L'_{-n}$ for $n>0$. 
Following the notation used in \cite{0005036}, we denote by 
$L^X_n$ the Virasoro generators
associated with the world sheet field $X$ 
and by $L'_n$ the
Virasoro generators associated with the other matter fields in the
boundary CFT. To the extent possible, we shall try to label the states
in terms of $L^X_{-n}$ instead of $\alpha^X_{-n}$.

\begin{figure}[!ht]
\leavevmode
\begin{center}
\epsfbox{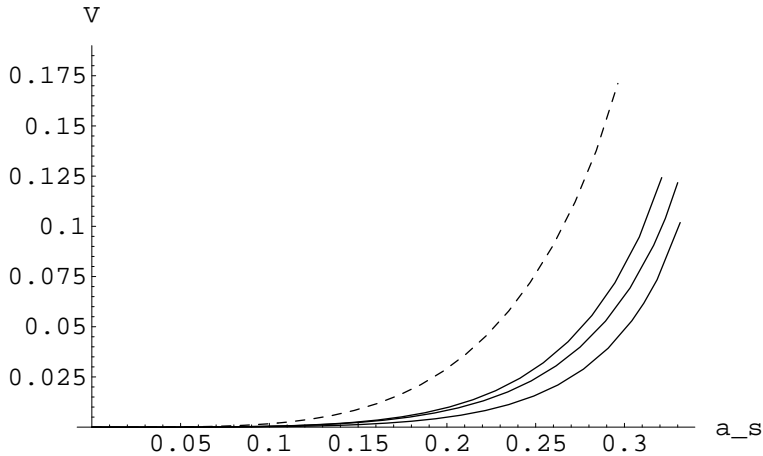}
\end{center}
\caption[]{\small The various effective potentials  $V\equiv(2\pi^2 
\VV^{M}(a_s))$ for 
the marginal parameter $a_s$ on
the $M$-branch. The dashed curve represents the level (1,2) approximation,
the successively flatter curves represent the level (2,4), (3,6) and (4,8)
approximations. Note that each potential 
has a different domain of definition.}
\label{f2}
\end{figure}

We can now easily construct the list of string fields appearing
at the next few levels. Let us begin with level two fields.
Since these are automatically twist even, they must also be
even under $X\to -X$. This means that we can get the usual
(Virasoro and ghost-current) descendents of the tachyon field 
but cannot get descendents
of the marginal field, since such descendents would be odd under
$X\to - X$.  As there is 
no new primary state in the CFT involving $X$ at this level,
the list of level 2 states in the string field is
\be
\label{sft2}
|\Phi^{(2)}\rangle = \Bigl( u_0 c_{-1}b_{-1} + v_0 L_{-2}^X 
+ w_0 L_{-2}'  \Bigr)\, c_1  
|0\rangle \,.  
\ee

Let us now continue to level three. Since all fields here are
twist odd, they must also be odd under $X\to -X$. We can therefore
allow all states obtained as ghost-current or Virasoro 
descendents of the level one state
$\alpha_{-1}^X c_1 |0\rangle$. One readily verifies that this
exhausts the list of possible states,\footnote{One can count this
as the number of level 3 ghost number 1 states built with 
$\alpha^X_{-n}$, $b_{-n}$, $c_{-n}$ and $L'_{-n}$ oscillators
which must have an odd number of 
$\alpha_{-n}^X$ oscillators.} and gives a total of
four level three fields:
\be
\label{sft3} 
|\Phi^{(3)}\rangle = \Bigl( s c_{-1}b_{-1} + \ron L_{-2}^X + \rtw L_{-2}' +
y L_{-1}^X L_{-1}^X \Bigr)\,
|\varphi_a\rangle\,,\quad |\varphi_a\rangle = \alpha_{-1}^X c_1
|0\rangle \,.
\ee
Finally we proceed to the list of fields at level
four. Being twist even, all states that
are ghost current or Virasoro descendents of the tachyon must be
included.
These give a total of ten states. There is one more twist even state at
this level, $-$
a level four primary of CFT($X$)
\be \label{defp4}
|p_4\rangle = \Bigl(  \alpha_{-3}^X \alpha_{-1}^X -  {3\over 4}
(\alpha_{-2}^X)^2
- {1\over 2} (\alpha_{-1}^X)^4 \Bigr) \, c_1 |0\rangle\, . 
\ee
Therefore the level four string field is given as
\begin{figure}[!ht]
\leavevmode
\begin{center}
\epsfbox{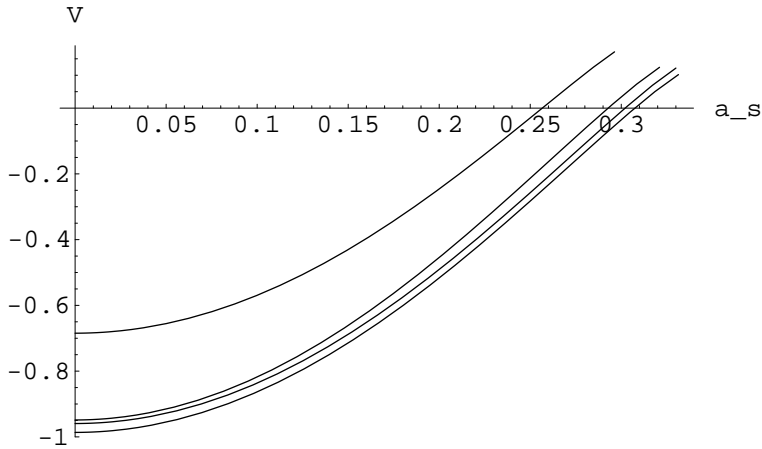}
\end{center}
\caption[]{\small The various effective potentials  $V\equiv(2\pi^2 
\VV^{V}(a_s))$ for 
the marginal parameter $a_s$ on
the $V$-branch. The top curve represents the level (1,2) approximation,
the successively lower curves represent the level (2,4), (3,6) and 
(4,8) approximations.} \label{f3}
\end{figure}
\ben
\label{sft3a}
|\Phi^{(4)}\rangle &=& g |p_4\rangle + 
\Bigl( \aon L_{-4}^X + \atw L_{-4}'  + \bon L_{-2}^X
L_{-2}^X + \btw L_{-2}' L_{-2}' + \bth L_{-2}' L_{-2}^X \cr\cr
&& + c\, c_{-3} b_{-1} + d\, b_{-3} c_{-1}  + e \,b_{-2} c_{-2}  + 
(\fon L_{-2}^X  + \ftw L_{-2}') c_{-1} b_{-1}  \Bigr)\, c_{1} |0\rangle\,.
\een
\begin{table}
\begin{center}\def\st{\vrule height 3ex width 0ex}
\begin{tabular}{|c|c|c|c|c|c|} \hline

Level & $\overline a_s$  & $\alpha_4^M$ & $\alpha_2^V$ &
$2 \pi^2 \VV^M({8\over27})$ & $2 \pi^2 \VV^M(\overline a_s)$ 
\st\\[1ex] \hline \hline

$(1,2)$ & 0.296296 & 0.843752 & 0.592593 & 0.1712 & 0.1712 \st\\[1ex]
\hline

$(2,4)$ & 0.321374 & 0.200234 & 0.672892 & 0.0743 & 0.1254
\st\\[1ex]
\hline

$(3,6)$ & 0.330107 & 0.200234 & 0.631329 & 0.0605 & 0.1221 \st\\[1ex]
\hline

$(4,8)$ & 0.331428 & 0.096999 & 0.633432 & 0.0444 & 0.1020 \st\\[1ex] \hline

\end{tabular}
\end{center}
\caption{We show the variation of various quatities as a function
of the level of the calculation. Here $\overline a_s$ denotes the maximal
value possible for the string field marginal parameter. 
The coefficient $\alpha_4^M$ 
defines the leading quartic term in $a_s$ in 
the effective potential on the $M$-branch.
The coefficient $\alpha_2^V$ 
defines the leading quadratic term in 
$a_s$ in the effective potential on the
$V$-branch. We also show the value of the potential, 
normalized in units of the tension 
of the brane, for the maximal value ($\overline a_s$) of $|a_s|$ 
at level
(1,2), and
for the end of the range at each level.}
\label{enable} 
\end{table}

We can now compute the potential $\VV$ at various levels of approximation
by
standard procedure. The results of this computation  are given in appendix
\ref{a1}. At each level, we can determine the effective potential for the
Wilson line in the $M$-branch by (numerically) eliminating all the other
fields by their
equations of motion. These results have been 
shown\footnote{Due to the $a_s\to -a_s$
symmetry of the potentials, we have displayed them only for positive
$a_s$.} 
in Fig.~\ref{f2}. We see
from this figure that the effective potential becomes flatter as we
increase the level of approximation. This can also be verified by
computing the coefficient $\alpha_4^M$ of the $a_s^4$ term in the
expression for the effective potential 
(see, for example, \refb{sas}), which has been listed in 
table \ref{enable} and is seen to decrease as we increase the level of
approximation. These results for $\alpha_4^M$ are in agreement 
with those
of ref.\cite{0001201}. In
each case we find a maximum value $\overline a_s$ of $|a_s|$ 
beyond which
the effective potential ceases to exist. These values have also been listed
in
table \ref{enable} and are  seen to converge rapidly to 
about 0.33. 
The
same procedure can be carried out to determine the effective potential in
the $V$-branch (by choosing different
initial data for obtaining the solution). These results have been shown in
Fig.\ref{f3}. As we see from this figure, the effective potential on the
$V$-branch does not become flat as we increase the level of approximation.
A quantitative measure of this is the coefficient $\alpha_2^V$ 
of the
$a_s^2$ term in the potential (see, for example, \refb{sbas}); 
as seen from table 1,
it converges to a finite value as we increase the level of approximation.

\sectiono{Tachyonic Lump Solution near Marginality}
\label{s3}

The setup here is that of Ref.~\cite{0005036}, namely, we consider a
D-brane with one of its spatial dimensions wrapped
around a circle of radius $R$. The object of interest is
the potential for string field modes that are either space-time
constants or  carry momentum along this circle. 
This problem
was indeed analysed in ref.\cite{0005036} for various values of
$R>1$.  Our
interest here, however,  is the $R\to 1$ limit where the
tachyonic mode $t_1$ carrying unit momentum along the circle becomes
marginal. Giving a vev to this mode can be shown to be equivalent to
giving a vev to a Wilson line~\cite{9902105}. Thus we expect this
string field tachyon condensation problem
to be related to the string field Wilson line problem  discussed in the
previous section. As we shall see by direct examination of the
corresponding string field potentials, 
this is indeed the case. We shall also be able to gain some additional
information about the Wilson line problem by exploiting this
equivalence.

In the same spirit as for the case 
of the marginal field, we  introduce radius
dependent effective potentials $\VVV (t_1; R)$ for 
$t_1$. We take the full string field potential
to a given approximation, and for fixed values
of $R$ and $t_1$  eliminate all other variables by using their equations
of motion. Choosing between the marginal\footnote{Note 
that the deformation associated with $t_1$ is no longer marginal for $R\ne
1$. But we shall continue to refer to this branch as the marginal branch.} 
or vacuum solution branches,
this defines, for the fixed chosen value of
$R$, the  effective potentials $\VVV^M (t_1; R)$ and 
$\VVV^V (t_1; R)$ for $t_1$. Except for the lowest level 
approximation, the effective potentials are calculated only
numerically, using the analytic expressions for the full
string field potential.

We also attempt here to estimate the vev of $t_1$ 
which at the 
radius $R=1$ leads to the
formation of the lump. Via the identification $t_1 \leftrightarrow
\sqrt{2} a_s$ to be established below, we obtain an estimate for
the vev of the string field parameter $a_s$
leading to the formation of the lump. 
On the other hand,
we can describe the formation of the lump
at $R=1$ as a marginal deformation of the boundary CFT. Let us denote by
$a_c$ the parameter labelling this marginal deformation,
normalised so that in the action of the deformed CFT, it multiplies $\int
dx \varphi(x)$,  with the marginal operator $\varphi$ normalized so that
$\langle\varphi|\varphi\rangle=1$. Then the value of $a_c$, which
corresponds to the formation of the lump solution at $R=1$, is
$a_c= \pm {1\over 2\sqrt 2}$\cite{RECK,9902105}. 
These pieces of information will be used in
section~\ref{s4} to investigate the functional
relationship between $a_s$ and $a_c$.

\subsection{Tachyon potential for D-brane on a circle at level (1,2)}

\label{tp12}

Since we 
will be particularly interested in the case $R\to 1$ where
the first tachyon harmonic $t_1$
becomes marginal, we will measure level as if the radius equals one.  
As in ref.\cite{0005036}, we shall consider string field configurations
which are even under twist, {\it and} the $X\to -X$ transformation. At
level one the string
field now includes the zero mode $t_0$ of the tachyon and 
its first harmonic $t_1$:
\be
\label{sft1}
|\wt\Phi^{(1)}\rangle = \Bigl( t_0  + t_1 \cos  
\Bigl({X(0)\over R}\Bigr) 
 \Bigr)\, c_1\, |0\rangle \,. 
\ee
The potential, given in eqns.~(3.5) and (3.6) of 
\cite{0005036} is 
\be
\label{vtachfirst}
\VVV (t_0, t_1; R) = -{1\over 2} t_0^2 
+ {1\over 3} K^3 t_0^3 - {1\over 4} (1 - {1\over
R^2}) t_1^2 + {1\over 2} K^{3 - 2/R^2} t_0 t_1^2 \,.
\ee
When $R=1$ the potential simplifies to 
\be
\label{vtachfi}
\VVV (t_0, t_1; R=1) = -{1\over 2} t_0^2 
+ {1\over 3} K^3 t_0^3 + {1\over 2 }\,K t_0 t_1^2 \, \,,
\ee
where $t_1$ now clearly represents a massless state. 
Comparing this with \refb{fgh} we see that 
the two potentials agree if we identify
\be
\label{ident}
t_1 \leftrightarrow \sqrt{2} a_s \,.
\ee
This factor of $\sqrt 2$ is simply a reflection of the fact that the
state $\cos(X(0)/R)|0\rangle$ has norm $(1/\sqrt 2)$, whereas
$\alpha^X_{-1}|0\rangle$ has unit norm. As we shall see in the next
subsection, this identification will be possible to implement
to higher level, as expected from the CFT argument~\cite{9902105}.

The tachyon case, however, lends itself to an
interesting analysis for $R$ slightly above one.
Again, we can integrate  $t_0$ using its field equation to
find an effective potential for the field $t_1$.
The quadratic equation for $t_0$ 
only has real solutions when 
\be \label{t1maxr}
|t_1| \leq \overline t_1 (R) \equiv       
{8\sqrt{2}\over 27} \,  K^{-1 +{1/ R^2}}\,.
\ee
Note that the maximum possible value $\overline t_1$ of $t_1$ 
becomes larger as the
radius is decreased towards the 
value $R=1$. The critical
value  $\overline t_1$ 
at $R=1$, as required, is in agreement with \refb{cra}
given
the identification \refb{ident}.
The resulting $M$-branch 
effective potential $\VVV^M (t_1; R)$ for
the 
$t_1$ field is a bit complicated and the explicit form is not
very illuminating, 
but it has a local minimum at:
\be \label{elmin}
t_1 = \pm K^{-3 + {2/ R^2}} \,\sqrt{1 - 1/R^2}\,\,\sqrt{ 
1 - {1\over 2} K^{2/R^2} (1 - {1\over R^2}) } \equiv \pm t_1^{(0)}(R)\, .
\ee
This describes the one lump solution for a given value of $R$. This value
of $t_1^{(0)}(R)$, of course, agrees with the analysis of
ref.\cite{0005036}. Here 
we would like to use $t_1^{(0)}(R)$ to 
estimate the vev of $t_1$ leading to the one lump solution at $R=1$,
namely, at marginality. The $R\to 1$ limit of $t_1^{(0)}(R)$, however,
simply vanishes!  The
source of this problem can be traced to the following facts. The effective
potential for $t_1$ has two parts: the term quadratic in $t_1$ which
vanishes identically at $R=1$, and the higher order terms which do not
vanish identically at a given level of approximation, but are expected to
vanish when we compute the potential exactly. Due to this, at {\it any}
given level of approximation the minimum of $\VVV^M (t_1;R)$ will
approach the $t_1=0$ point as $R\to 1$. We shall try to get around this
problem by working at a fixed value of $R$ close to 1 ({\it e.g.}
$\sqrt{1.1}$), and then increasing the level of approximation to
determine the point that
$t_1^{(0)}$ approaches for this fixed value of $R$.

\subsection{Tachyon potential for D-brane on a circle at higher level}
\label{hla}

We now include fields at higher levels. Keeping only states which are
even under twist and the $X\to -X$ transformations
as in \cite{0005036}, 
we get the following
set of states at level two:
\be
\label{sft2b}
|\wt\Phi^{(2)}\rangle = 
\Bigl( u_0 c_{-1}b_{-1} + v_0 L_{-2}^X 
+ w_0 L_{-2}'  \Bigr)\, c_1
|0\rangle \,, 
\ee
Note that this is identical to the list of states
we generated in \refb{sft2} for the Wilson line problem.

At level three we have, 
\be
\label{sft3b}
|\wt\Phi^{(3)}\rangle = 
\Bigl( u_1 c_{-1}b_{-1} + v_1 L_{-2}^X + w_1
L_{-2}'
+
z_1 L_{-1}^X L_{-1}^X \Bigr)\,
|\varphi_t\rangle\,,\quad |\varphi_t\rangle = \cos\bigg({X(0)\over
R}\bigg) c_1 |0\rangle \,.
\ee
At level four the fields to be included are:
\ben
\label{sft4b}
|\wt\Phi^{(4)}\rangle &=& 
\wt g |p_4\rangle + t_2 |\chi\rangle + 
\Bigl( \aon L_{-4}^X + \atw L_{-4}'  + \bon L_{-2}^X
L_{-2}^X + \btw L_{-2}' L_{-2}' + \bth L_{-2}' L_{-2}^X \cr\cr
&& + c\, c_{-3} b_{-1} + d\, b_{-3} c_{-1}  + e \,b_{-2} c_{-2}  + 
(\fon L_{-2}^X  + \ftw L_{-2}') c_{-1} b_{-1}  \Bigr)\, c_{1} |0\rangle\,,
\een
where $|p_4\rangle$ has been defined in eq.\refb{defp4}, and 
\be \label{esmode}
|\chi\rangle = \cos({2 X(0)\over R}) |0\rangle\, .
\ee
The string field theory potential involving 
these fields has been computed
and given in appendix \ref{a2}.

\subsection{Relating marginal tachyons to Wilson lines}

As we have mentioned several times, 
at $R=1$ 
the CFT describing the D-brane marginal tachyon dynamics can be mapped
to a Wilson line CFT problem.  We have already seen that at level
(1,2) this identification is indeed realized by setting 
$t_1 \leftrightarrow \sqrt{2}
a_s$ as indicated in \refb{ident}. We shall now explain how this
identification can be extended to show that the $R=1$ tachyon
problem is completely isomorphic to the Wilson line problem
to level (4,8). It follows from this that the effective potential
$\VV^M (a_s)$ for the Wilson line parameter and
$\VVV^M (t_1;R)$ for the first tachyon harmonic are 
related as
\be
\label{relpot}
\VV^M (a_s) = \VVV^M ( t_1 = \sqrt{2} a_s ; R=1)\,.
\ee

We begin by arguing equivalence of the types of terms
present in the complete string
field potentials. We shall denote by
$\VV^{(M,N)}$ the level $(M,N)$ approximation to $\VV(\Phi)$, by
$\VV_{mm}$ the quadratic term in the potential for level $m$ fields, and
by $\VV_{mnp}$ the cubic term in the potential coupling a level $m$, a
level $n$ and a level $p$ field. 
As we did before, the potential terms $\VVV$ 
for the tachyon lump case  are distinguished from the
Wilson line potential terms by the tilde. We now claim that
the total level of all the terms entering a term in the potential  must be
even. For the case of the tachyon lump this follows 
readily 
by invariance of the string field action 
under  the translation $X\to X+\pi
R$; the odd level fields carry odd unit of momentum and hence are odd under
this transformation  whereas 
even level states carry even unit of momentum and so are even under this
transformation. For the case of the Wilson line this follows from
$X\to -X$ symmetry under which the odd level fields are odd, and
the even level fields are even. 
These symmetries are eventually
responsible for the $t_1\to -t_1$ and $a_s\to -a_s$ symmetries of the
corresponding effective potentials.

We list below, for convenience, 
the terms that appear in the
string field potential relevant for the study 
of the Wilson line when we include fields up
to a given level
(while keeping the total level below 8):
\ben \label{esubeqn} 
\hbox{level zero} &:& \VV_{00} \,,  \VV_{000}\nonumber \\
\hbox{level one} &:&  \VV_{11}\, ,\, \VV_{011} \nonumber \\
\hbox{level two} &:& \VV_{22} \,,\, \VV_{002} \,,\, \VV_{112}\,,
\VV_{022}\,,\, 
\VV_{222} \,,\nonumber \\
\hbox{level three} &:& \VV_{33} \,,\, \VV_{013} \,,\,
\VV_{033} \,,\, \VV_{123} \,,\, \VV_{233}\nonumber \\
\hbox{level four} &:&  \VV_{44}\,,\,\VV_{004} \,,\, \VV_{114} \,,\,
\VV_{024} \,,\, \VV_{224} \,,\, \VV_{044} \,,\, \VV_{134} 
\een
For each level $\ell$ the list of terms up to 
and including
those in the appropriate line give all interactions involving
fields with level less than or equal to $\ell$.
A similar list with $\VV$ replaced by $\VVV$ appears for the string field
potential relevant for the study of the lump solution on a circle.

With the identification $t_1 \leftrightarrow \sqrt{2} a_s$ we have
already guaranteed that the terms listed in the first two lines above give 
exactly the same potentials as their tilde versions.  
Indeed, with
$\varphi_a$ and $\varphi_t$ used to denote the Wilson and
tachyon unit momentum marginal operators (see \refb{sft3} and
\refb{sft3b}) the agreement is the result of the following equality of
correlators involving the Virasoro and ghost current primaries
$\varphi_a$ and $\varphi_t$:
\be
\label{bequ}
\langle {\varphi_a\over \sqrt{2}}\,,\, {\varphi_a\over \sqrt{2}}\,, t\,
\rangle =
\langle
\varphi_t\, ,
\varphi_t\, , t\, \rangle \,.
\ee
In this equation $t$ denotes the CFT vertex 
operator for the zero-momentum
tachyon $c_1|0\rangle$.

Let us
now consider the third line in \refb{esubeqn}, that is include the new
terms that involve level two fields (and fields with lower level). Note
again that the level two fields in both cases (eqns.~\refb{sft2} and
\refb{sft2b}) are identical and in fact we have used the same labels for
them. It follows that all couplings involving level zero and level two
fields in both string field potentials are identical. The only question
is whether the $\VV_{112}$ terms are the same as the $\VVV_{112}$ terms. 
In other words,  does the equality 
\be
\label{bequ2}
\langle {\varphi_a\over \sqrt{2}}\,,\, {\varphi_a\over \sqrt{2}}\,,
\Phi^{(2)}\,
\rangle = \langle \varphi_t\, ,
\varphi_t\, , \wt\Phi^{(2)}\, \rangle 
\ee
hold for any of the three level two fields in $\Phi^{(2)}$ and $\wt
\Phi^{(2)}$ ? 
It does. Indeed, all fields in $\Phi^{(2)}$ ($\wt
\Phi^{(2)}$) 
are either Virasoro or $U(1)$
ghost-current descendents of the level zero tachyon. 
Since both $\varphi_a$ and $\varphi_t$ have the same dimension
and ghost number, the
equality \refb{bequ2} follows from \refb{bequ}.

Let us
now consider the fourth line in \refb{esubeqn}, that is include the new
terms that involve level three fields (and fields with lower level).
Note the complete isomorphism manifest from equations 
\refb{sft3} and \refb{sft3b}: all these fields are 
just Virasoro and ghost-current descendents of $\varphi_a$ and
$\varphi_t$ respectively. It follows from the normalizations
of $\varphi_a$ and
$\varphi_t$ that $\VV_{33}$  and $\VVV_{33}$  
terms will match if we identify
\be \label{l3idenf}
s = {u_1\over \sqrt 2}\, , \quad \ron = {v_1 \over \sqrt 2}\,, \quad \rtw
= {w_1\over \sqrt 2}\, , \quad y = {z_1\over \sqrt 2}\,.
\ee
Our remarks about descendents imply that the equality
of correlators defining $\VV_{013}$,
$\VV_{033}$, $\VV_{123}$, and  $\VV_{233}$, with their
tilde counterparts 
simply follow from \refb{bequ}.

Finally, let us
consider the fifth and final line in \refb{esubeqn}, giving
the new terms that involve level four fields (and fields with lower
level). We do this in two stages. Let us first consider the
common list of fields in \refb{sft3a} and \refb{sft4b}. 
These ten fields, denoted by the same labels,
are all Virasoro and ghost
current descendents of the zero momentum tachyon.
By the earlier arguments, all correlators involving these level
four fields and fields of level $\le 3$ 
will agree in the Wilson and tachyon string field
potentials. Now consider the remaining fields $g$ (multiplying 
$|p_4\rangle$)
in the Wilson case, and, $\wt g$ and $t_2$ (multiplying $|p_4\rangle$
and $|\chi\rangle$ respectively) 
in the tachyon lump case. All of these states are primaries. 
We now show that out of the two level four primaries in the
tachyon lump case, only one linear combination needs to be 
kept at $R=1$. 

For this purpose consider in the tachyon lump case the complete
list of primaries at all levels less than or equal to four appearing in
the study of $\VVV$: 
\be
\label{lprim}
\{ \, t\,, \, \varphi_t \,,\, p_4\, ,\,  \chi \, \}
\ee
We now split them as follows
\be
\label{lprima}
\Bigl\{ \, t\,, \, \varphi_t \,,\, \wt p_4 \equiv {1\over 2} 
(-p_4 + 9\chi)\,\Bigr\}  
\,, \qquad d \equiv  {1\over 2} (p_4 + 3\chi )\,.
\ee
In the first set we have included three vertex 
operators, and we
have separated out a fourth 
one called $d$  ($d$ stands for decoupled). 
We now claim that the kinetic terms in the string field
theory do not mix $d$ with any  operators in the first set.
For the first two operators this is trivially so, and for
the third it follows from $\langle p_4 |c_0 L_0 |p_4\rangle = 81/2$,
$\langle \chi |c_0 L_0 |\chi\rangle = 3/2$, 
and $\langle p_4 |c_0 L_0|\chi\rangle=0$. 
Moreover, $\wt p_4$
has been normalized such that 
$\langle \wt p_4 |c_0 L_0 |\wt p_4\rangle= \langle p_4 |c_0 L_0
|p_4\rangle$ and also satisfies the property that any three
point correlator in the first set involving one or two $\wt p_4$'s equals
the correlator with $\wt p_4$ replaced by 
$p_4$ and $\varphi_t$ replaced by $\varphi_a/\sqrt 2$. 
(This can be
checked by explicit computation.) 
Finally,
any three point correlator
involving two fields from the first set and the field $d$ vanishes.  This
is trivially so for $\langle t,t,d\rangle$ and less trivially so for
$\langle \varphi_t, \varphi_t, d\rangle$ and others. 

It follows from the above argument that if we rewrite the 
relevant part of the string field $|\wt\Phi^{(4)}\rangle$ 
making use of the new basis
\be 
\label{nbas}
\wt g\, |p_4\rangle  + t_2 \,|\chi\rangle = {1\over 6} (t_2 - 3\wt g)
\,|\wt p_4\rangle + {1\over 6} \,(t_2 + 9\wt g )\, | d\rangle\,,
\ee
the component field associated
to $|d\rangle$ will not acquire an expectation value as it has
no one point functions with other fields that do. In addition, the
interchangeability of 
$(p_4,\varphi_a/\sqrt 2)$ and $(\wt p_4,\varphi_t)$ 
in the relevant
computations indicates that the coefficient field of 
$|\wt p_4\rangle$ in $|\wt\Phi^{(4)}\rangle$  
should be identified with the field $g$, $-$ 
the coefficient of $|p_4\rangle$ 
in $|\Phi^{(4)}\rangle$. 
All in
all we have that the interactions involving level four fields will
agree upon the identifications
\be
g = {1\over 6} (t_2 - 3\wt g) \,, \quad \hbox{and}, \quad
0 =  {1\over 6} (t_2 + 9\wt g ) \,. 
\ee
In view of this and  previous identifications, our result for the
equivalence of the Wilson and tachyon lump string field
potentials (ommitting all common field variables that are simply
identified) reads 
\ben
\VV \Bigl(a_s,  \{ s, \ron, \rtw, y\}, g \Bigr) 
= \VVV \Bigl(t_1 = \sqrt{2} a_s,  \{ u_1, v_1, w_1, z_1\}= \sqrt{2}
 \{ s, \ron, \rtw, y\}, \nonumber\\ \wt g = - g/2,  t_2 = 9g/2 \,\,;  R=1
\Bigr)
\een
This is the main result of this subsection.  
Integrating out all fields
except
$t_1$ or $a_s$ yields the result quoted in \refb{relpot}.

\subsection{Estimating the string field marginal parameter for the lump}

Having noted at the end of 
subsection~\ref{tp12} that taking
the limit $R\to 1$ at any fixed level does not provide an
estimate for the vev of $t_1$ at the lump solution, we try
now taking a fixed value of $R$ near one, and examine how
the vev of $t_1$ at the lump solution varies as we increase
the level. We will take, somewhat arbitrarily, $R= \sqrt{1.1}$,
a value of $R$ reasonably close to one, but (hopefully) 
not so close
that
we would need a prohibitively high level computation to converge
to the expected values of $t_1$.

\begin{table}
\begin{center}\def\st{\vrule height 3ex width 0ex}
\begin{tabular}{|c|c|c|c|c|} \hline

Level & $\overline t_1$  & $t_1^{(0)}$ & $t_0^{(0)}$ &
$t_2^{(0)}$
\st\\[1ex] \hline \hline

$(1,2)$ & 0.4092 & $\pm$ 0.21307 & 0.03336 & --- \st\\[1ex]
\hline

$(2,4)$ & 0.4462 & $\pm$ 0.29707 & 0.06964 & ---
\st\\[1ex]
\hline

$(3,6)$ & 0.4598 & $\pm$ 0.31127 & 0.07693 & ---  \st\\[1ex]
\hline

$(4,8)$ & 0.4624 & $\pm$ 0.33625 & 0.09425 & $-0.0102$  \st\\[1ex] \hline

\end{tabular}
\end{center}
\caption{We show the variation of various quatities as a function
of the level of the calculation for $R=\sqrt{1.1}$. 
Here $\overline t_1$ denotes the maximal
value possible for the tachyon harmonic $t_1$. 
The next three columns give the values of the tachyon 
harmonics at 
the lump
solution of the equations of motion. 
Note that as the
level is increased, the vev of the nearly marginal
tachyon harmonic $t_1$ increases.}
\label{t2} 
\end{table}

\begin{figure}[!ht]
\leavevmode
\begin{center}
\epsfbox{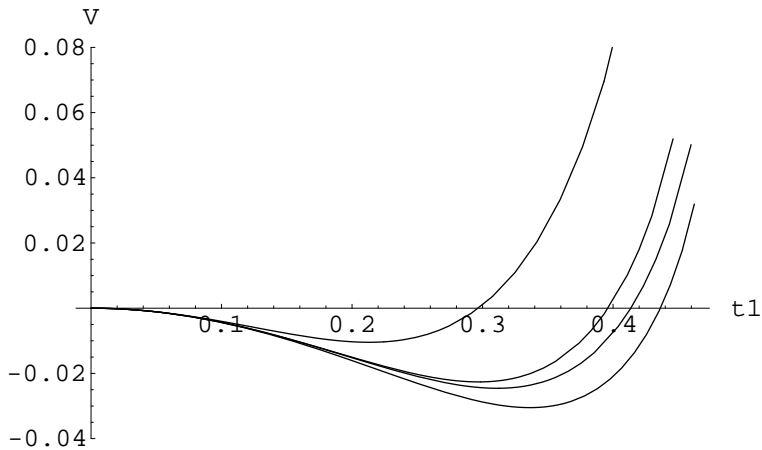}
\end{center}
\caption[]{\small The level (1,2), (2,4), (3, 6) and  (4,8) effective
potentials $V\equiv 2\pi^2\VVV^M (t_1;R)$ for  
$t_1$ when $R= \sqrt{1.1}$. As the level is increased
 the potentials
become deeper and the value of $t_1$ at the minimum larger.} \label{f5}
\end{figure}

Using the potential given in appendix \ref{a2}, and
setting $R= \sqrt{1.1}$, we can calculate the value of
the string field at the extremum of the potential representing
the lump at different levels of
approximation.
In table \ref{t2} we have given the 
values $t_0^{(0)}$, 
$t_1^{(0)}$ and $t_2^{(0)}$ of the tachyon harmonics at this extremum.
Further insight is obtained by consideration
of the effective potentials $\VVV^M (t_1; R=\sqrt{1.1})$ obtained 
at various approximation levels and shown in 
Fig.~\ref{f5}. Note that 
with increasing level,
the potentials become increasingly deep and the minimum is
attained for larger and larger values of $t_1$. Those
values 
$t_1^{(0)}$  for $t_1$ are the ones given on the table. 
At level (4,8), the value of $t_1$ at the minimum of
the potential is $\pm .336$.  Assuming that this is a good
approximation to $t_1^{(0)}$ for $R=1$, and using 
eq.\refb{ident}, we see 
that the lump solution at $R=1$ corresponds to $a_s=\pm {.336/ \sqrt{2}}  
=\pm .238$. This gives 
\be \label{elump} 
a_s \simeq\pm .238 \qquad \hbox{at} \qquad a_c=\pm{1\over 2\sqrt 2}\,. 
\ee
Due to the reasons mentioned at the end of 
subsection~\ref{tp12}, 
however, this value of $t_1^{(0)}$ might not be a very accurate
result, since even at level (4,8)  
$\VV^M(a_s)=\VVV^M(\sqrt 2a_s, R=1)$ 
receives an
appreciable contribution, and hence causes a significant distortion of 
the potential at $R=\sqrt{1.1}$. Indeed, the pattern in table~\ref{t2}
suggests that at least at this radius, we are underestimating the
value of $t_1^{(0)}$.

\sectiono{Matching CFT and SFT Marginal Parameters} \label{s4}

In this section we shall try to interpret our results of section 
\ref{s2}. In particular we shall be addressing the question: what does it
mean to have a finite cut-off $\overline a_s$ on $a_s$ beyond which the
string field
theory calculation of the effective potential breaks down?

Clearly the most important question here is: what gauge field vev
does the point $a_s=
\overline a_s$ correspond to?  If it corresponds to infinite value of
this gauge field then our
results would imply that in string field theory a finite range of the
string field covers the full range of values of the 
marginal deformation parameter 
in conformal field theory. On the other hand if $a_s=
\overline a_s$ corresponds to a finite value of the gauge field vev, then
it
would mean that formulated around a given background, the string field
theory only covers a finite subset of the full CFT moduli space.

In order to be able to resolve this issue, one needs to know the
relationship between the string field theory parameter $a_s$ and the
gauge field vev $a_c$ in the Born-Infeld action.
In general the parameters $a_s$ and $a_c$ are related by a
function:\footnote{Some aspect of this relationship 
has been discussed recently in ref.\cite{0005085}.}
\be \label{efunc}
a_s = f(a_c)\, .
\ee
If $|\varphi\rangle$ denotes the normalized dimension one primary
state
representing the marginal direction, and $\varphi$ denotes the
corresponding vertex operator, then we take $a_s$ to be the coefficient of
the state $c_1|\varphi\rangle$ in the expansion of the string field, and
$a_c$ to be coefficient of $\int dx \varphi(x)$ to be added to the CFT
action in order to construct the marginally deformed CFT. With this
normalization convention, for small $a_c$, $a_s\simeq a_c$~\cite{BACK}.
This gives
$f(a_c)\simeq a_c$ for small $a_c$.

Our interest lies in studying the behaviour of $f(a_c)$ for large $a_c$.
In particular, we want to explore 
the possibility that $a_s=\overline
a_s$
corresponds to
$a_c=\infty$. It is easy to construct functions which approach a finite
value $\overline a_s$  for large $a_c$. 
An example of a function of this type is: 
\be
\label{conj}
f(a_c) = \overline a_s \tanh \,( \,{a_c\over\overline a_s}\,\,)\, .
\ee
This would have the requisite properties
$f(a_c) \approx a_c$ for $a_c$ small, and  $f (\infty) = \overline a_s$.
Using the level (4,8) value of $\overline a_s$ ($=.331$), eq.\refb{conj}
predicts:
\be \label{epredict}
f(a_c={1\over 2\sqrt 2}) = .331 \tanh{1\over .331 \times 2 \sqrt 2} =
.261\, .
\ee
This is in fair agreement with eqs.\refb{elump}. 
Indeed, as remarked below that equation, \refb{elump} is probably an
underestimate of the actual value of $a_s$ for $a_c={1\over 2\sqrt 2}$.

Although this analysis seems to indicate that eq.\refb{conj} gives a
fairly
accurate description of the relationship between $a_s$ and $a_c$, there
is also counterevidence to this conjectured relationship.
For this we again turn to the potential $\VVV^M(t_1;R)$
in the tachyonic lump problem.
According to the analysis of ref.\cite{9902105}, for $R>1$ but close to 
1, the effective potential is periodic in $a_c$ with periodicity $1/\sqrt
2$. This means that the potential should have an infinite 
number of oscillations in the range $0\le a_c<\infty$, 
and in particular have a maximum at
$a_c={1\over \sqrt 2}$. According to eq.\refb{conj} this corresponds to
the point $a_s=.322$, {\it i.e.} $t_1=\sqrt{2} \times .322 = .455$.
Examining Fig.~\ref{f5} we find no evidence for
a maximum of $\VVV^M(t_1;R)$ near $t_1\sim .455$,
nor any oscillation of
the
potential. This seems to indicate that eq.\refb{conj} does not quite
represent the correct relation between
$a_s$ and $a_c$, and that $a_s=\overline a_s$ 
may correspond to a value of $a_c$
below ${1\over \sqrt 2}$. We should
recall, however,  that $\VV^M(a_s)=\VVV^M(\sqrt 2 a_s, R=1)$ 
computed 
at level
(4,8) has a large
slope
at $a_s=.322$ (see Fig.~\ref{f2}), and this might destroy a
potential maximum in
$\VVV^M(\sqrt{2}a_s ; R)$ at this point. Thus in absence of better
numerical results, we are unable to
decide whether $a_s=\overline a_s$ corresponds to infinite or finite
$a_c$.

\sectiono{Concluding Remarks} \label{s5}

The finite range of definition of the effective
potential for the string field marginal parameter $a_s$
in the (marginal) $M$-branch arose because beyond
some limiting value for $a_s$ field equations 
for other string fields could not be solved at all
or solutions would fail to be continuous functions of $a_s$.
More concretely, 
we saw that at the critical
value for $a_s$ the $M$-branch and the $V$-branch, associated
with the stable tachyonic vacuum merged.

Actually, even the tachyon effective potential
appears to have a finite range of definition
in the level expansion \cite{KS,0002237} with the
stable minimum 
well inside this range. In this
case the limiting values appear as points beyond
which some massive field equations either fail to have
solutions or fail to give solutions that are
continuous functions of the tachyon. 
The physical interpretation of this 
finite range is not clear; in particular,
the lower limiting value
is precisely in the direction where the effective
potential of the tachyon is expected to be 
unbounded below. A complete understanding of the 
more familiar marginal
case discussed in this paper might help interpret 
the finite range of the tachyon effective potential. 

For the case of superstring field theory on a BPS D-brane,
the non-BPS D-brane, or the D-brane anti-D-brane pair, 
if string field marginal parameters have finite ranges
it will be through an effect technically similar to
that of the tachyon effective potential. Given that in 
such superstring field theories the fields in the
GSO odd sector 
acquire no expectation values, unfamiliar
branches of solutions associated to massive fields in the
GSO even sector would have to limit the domain of 
definition of the effective potential.  

It certainly appears that more 
accurate calculations could give significant insight into the  
questions raised in this paper.
A proper understanding of the description of marginal
operators in both bosonic string theory and superstring theory
promises to deepen considerably our understanding of 
the way non-perturbative physics is encoded in string field
theory. 

\bigskip

{\bf Acknowledgement}: We are grateful to W. Taylor
for detailed discussions on various aspects of this work
and for comparing with us his unpublished results.
We would also like to thank L. Rastelli for useful discussions. 
A.S. would like to thank the New High Energy Theory Center at Rutgers
University, Center for Theoretical Physics at MIT and Erwin Schroedinger
Institute, Vienna for hospitality during the course of the work.
The work of B.Z. was supported in part
by DOE contract \#DE-FC02-94ER40818.

\appendix

\sectiono{SFT Potential for Study of Wilson Line}  
\label{a1}

In this appendix we shall derive the string field theory potential
$\VV(\Phi)$ (with normalization as defined in the text) relevant for
studying the effective potential for
constant gauge field configuration.
The expansion of the string field has been described
in section \ref{s2}. We shall denote by
$\VV^{(M,N)}$ the level $(M,N)$ approximation to $\VV(\Phi)$, by
$\VV_{mm}$ the quadratic term in the potential for level $m$ fields, and
by $\VV_{mnp}$ the cubic term in the potential coupling a level $m$, a
level $n$ and a level $p$ field. As discussed in the text, using twist
symmetry of the action, it 
is easy to see that the total level of all the fields
entering a term in the potential  must be even. We then have: 
\ben \label{esubeq1} 
\VV^{(0,0)} &=& \VV_{00} + \VV_{000}\nonumber \\
\VV^{(1,2)} &=& \VV^{(0,0)} + \VV_{11} + \VV_{011} \nonumber \\
\VV^{(2,4)} &=& \VV^{(1,2)} + \VV_{22} + \VV_{002} + \VV_{112} +\VV_{022}
\nonumber \\
\VV^{(3,6)} &=& \VV^{(2,4)} + \VV_{33} + \VV_{013} + \VV_{222} + \VV_{033}
+ \VV_{123} \nonumber \\
\VV^{(4,8)} &=& \VV^{(3,6)} + \VV_{44} + \VV_{004} + \VV_{114} +
\VV_{024} + \VV_{233} + \VV_{224} + \VV_{044} + \VV_{134} \nonumber\\
\een
We let $K=3 \sqrt{3} / 4$ as in the text. 
Explicit computation gives the following expressions for $\VV_{mm}$ and
$\VV_{mnp}$:
\ben \label{emaineq1}
\VV_{00} &=& -{1\over 2} t_0^2 \nonumber \\
\VV_{11} &=& 0 \nonumber \\
\VV_{22} &=& -{1\over 2} u_0^2 + {1\over 4} v_0^2 + {25\over 4} w_0^2
\nonumber \\
\VV_{33} &=& - s^2  + {9\over 2} \ron^2 + {25\over 2} \rtw^2 + 12 y^2 
+ 12 \ron y \nonumber \\
\VV_{44} &=& {15\over 2} \aon^2 + {375\over2} \atw^2 + 9 \aon \bon +
{27\over
4}
\bon^2 + 225 \atw \btw + {2475\over 4} \btw^2 + {75\over 8} \bth^2 + 3 c d 
\nonumber \\
&& -
{3\over 2} e^2 - {3\over 4} \fon^2 -{75\over 4} \ftw^2 + {81\over 4} g^2
\nonumber \\
\VV_{000} &=& {1\over 3} K^3 t_0^3 \nonumber \\
\VV_{011} &=& K t_0 a_s^2 \nonumber \\
\VV_{002} &=& {1\over 32} K t_0^2 (22 u_0 - 5 v_0 - 125 w_0) \nonumber \\
\VV_{112} &=& {1\over 32} K^{-1} a_s^2 
(22 u_0 + 27 v_0 - 125 w_0)
\nonumber \\
\VV_{013} &=& {1\over 16} K^{-1} t_0 a_s \Big( 22 s - 21 \ron -
125 \rtw
-12 y\Big)\nonumber \\
\VV_{022} &=& {1\over 1024} K^{-1} t_0 (228 u_0^2 - 220 u_0 v_0 + 537
v_0^2
- 5500 u_0 w_0 + 1250 v_0 w_0 + 28425 w_0^2) \nonumber \\
\VV_{004} &=& {1\over 1024} K^{-1} t_0^2 (540 \aon + 13500 \atw + 459 \bon +
26475 \btw +
625 \bth \nonumber \\
&& - 320 c + 960 d + 228 e - 110 \fon - 2750 \ftw)
\nonumber \\[3ex]
\VV_{222}
& = & \displaystyle{
K \left\{ {1 \over 144} u_0^3 + {8321 \over 93312} 
v_0^3 - {219775 \over 10368} w_0^3  
- {95 \over 7776} u_0^2 
\left( v_0 + 25 w_0 \right) \right. } \nonumber \\
& & \displaystyle{
+ {1969 \over 15552} u_0 v_0^2 + 
{104225 \over 15552} u_0 w_0^2 - 
{22375 \over 31104} v_0^2 w_0} 
\displaystyle{\left. 
- {47375 \over 31104} v_0 w_0^2 
+ {6875 \over 23328} u_0 v_0 w_0 \right\} } \nonumber \\[3ex]
\VV_{033} &=& {1\over 1728} K^{-1} t_0
(228 s^2 - 924 s \ron + 9657 \ron^2 - 5500 s \rtw + 5250 \ron \rtw + 
    28425 \rtw^2 \nonumber \\
&& \qquad - 528 s y + 31224 \ron y  + 
    3000 \rtw y + 30864 y^2)
\nonumber
\\[3ex]
\nonumber \\
\VV_{123} &=& {1\over 864} K^{-1}
a_s (228 u_0 s + 594 s v_0 - 462 u_0 \ron + 969 v_0 \ron - 2750 s w_0
+ 
      2625 \ron w_0 \nonumber \\
&& - 2750 u_0 \rtw - 3375 v_0 \rtw 
+ 28425 w_0 \rtw - 264 u_0 y - 
      324 v_0 y + 1500 w_0 y)
\nonumber \\[3ex]
\VV_{114} &=& {1\over 1728} K^{-1}
a_s^2 (-868 \aon + 13500 \atw - 885 \bon + 26475 \btw - 3375 \bth - 320 c
\nonumber \\
&& \qquad +
960 d + 
      228 e + 594 \fon - 2750 \ftw +3072 g)
\nonumber \\[3ex]
\VV_{024} &=& {1\over 16384} K^{-3} t_0 \Big( u_0
( 11880 \aon + 297000 \atw + 10098 \bon+ 582450 \btw +
13750 \bth - 
      9600 c \nonumber \\
&& \qquad \qquad + 28800 d + 30616 e - 1140 \fon - 28500 \ftw)
\nonumber \\
&& \qquad + v_0 ( 
      7540 \aon - 67500 \atw - 23799 \bon - 132375 \btw -
67125 \bth + 
      1600 c \nonumber \\ 
&& \qquad \qquad - 4800 d - 1140 e + 11814 \fon + 13750 \ftw)
\nonumber \\
&& \qquad + w_0
(- 
      67500 \aon - 1431500 \atw - 57375 \bon - 6918975 \btw - 
      142125 \bth \nonumber \\
&& \qquad \qquad + 40000 c - 120000 d - 28500 e + 13750 \fon + 
      625350 \ftw)\Big)
\nonumber \\
\VV_{233} &=& {1 \over 18432} K^{-3}  
(648 u_0 s^2 + 2052 s^2 v_0 - 3192 u_0 s \ron + 14212 s v_0 \ron + 
    70818 u_0 \ron^2 \nonumber \\ 
&& \qquad + 16257 v_0 \ron^2 - 9500 s^2 w_0 +
38500 s \ron w_0
- 
    402375 \ron^2 w_0 - 19000 u_0 s \rtw \nonumber \\
&& \qquad - 49500 s v_0 \rtw +
38500 u_0 \ron \rtw - 
    80750 v_0 \ron \rtw + 416900 s w_0 \rtw \nonumber 
\\
&& \qquad - 397950 \ron w_0 \rtw +
208450 u_0 \rtw^2 + 
    255825 v_0 \rtw^2 - 1977975 w_0 \rtw^2 \nonumber \\
&& \qquad - 1824 u_0 s y -
4752 s v_0 y + 
    228976 u_0 \ron y - 58952 v_0 \ron y \nonumber \\
&& \qquad + 22000 s w_0 y
-
1301000 \ron w_0 y + 
    22000 u_0 \rtw y + 27000 v_0 \rtw y \nonumber \\
&& \qquad - 227400 w_0 \rtw y
+
226336 u_0 y^2 - 
    49904 v_0 y^2 - 1286000 w_0 y^2)
\nonumber \\[3ex]
\VV_{224} &=& {1\over 1048576} K^{-5} \Big(
u_0^2 (123120 \aon + 3078000 \atw + 104652 \bon 
+ 6036300 \btw + 
    142500 \bth \nonumber \\
&& \qquad - 103680 c + 311040 d +
1997584 e - 
    9720 \fon - 243000 \ftw) 
\nonumber \\
&& \qquad + u_0 v_0 ( 331760 \aon 
- 2970000 \atw - 
    1047156 \bon - 5824500 \btw - 2953500 \bth
\nonumber \\
&& \qquad
+ 96000 c - 
    288000 d - 306160 e + 244872 \fon 
+ 285000 \ftw) \nonumber \\
&& \qquad + v_0^2 (- 
    1254212 \aon + 7249500 \atw + 658131 \bon + 14217075 \btw - 
    3120375 \bth \nonumber \\
&& \qquad - 171840 c + 515520 d +
122436 e + 
    549186 \fon - 1476750 \ftw) \nonumber \\
&& \qquad + u_0 w_0 (- 2970000 \aon
- 
    62986000 \atw - 2524500 \bon - 304434900 \btw - 
    6253500 \bth \nonumber \\
&& \qquad + 2400000 c - 7200000 d - 7654000 e + 
    285000 \fon + 12961800 \ftw) \nonumber \\
&& \qquad + v_0 w_0 (- 1885000 \aon + 
    14315000 \atw + 5949750 \bon + 69189750 \btw + 
    30528450 \bth \nonumber \\
&& \qquad - 400000 c + 1200000 d + 285000 e - 
    2953500 \fon - 6253500 \ftw) \nonumber \\
&& \qquad + w_0^2 (15349500 \aon + 
    283692700 \atw + 13047075 \bon + 1777831875 \btw + 
    29669625 \bth \nonumber \\
&& \qquad - 9096000 c + 27288000 d +
6480900 e - 
    3126750 \fon - 130546350 \ftw)\Big)
\nonumber \\[3ex]
\VV_{044} &=&{1\over 1048576} K^{-5} t_0
(18846480 \aon^2 + 14580000 \aon \atw + 646122000 \atw^2 
+ 21655656 \aon \bon 
\nonumber \\
&& \qquad \qquad + 
      12393000 \atw \bon + 10551033 \bon^2 + 28593000 \aon \btw 
+ 1120943400 \atw \btw \nonumber \\
&& \qquad \qquad + 
      24304050 \bon \btw + 2269335225 \btw^2 
- 1885000 \aon \bth + 14315000 \atw \bth \nonumber \\
&& \qquad \qquad + 
      5949750 \bon \bth + 69189750 \btw \bth 
+ 15264225 \bth^2 - 345600 \aon c - 
      8640000 \atw c \nonumber \\
&& \qquad \qquad - 293760 \bon c - 16944000 \btw c - 400000
\bth c + 
      76800 c^2 + 1036800 \aon d \nonumber \\
&& \qquad \qquad + 25920000 \atw d + 881280 \bon
d + 
      50832000 \btw d + 1200000 \bth d + 1636352 c d \nonumber \\
&& \qquad \qquad + 691200
d^2 + 
      246240 \aon e + 6156000 \atw e + 209304 \bon e + 12072600 \btw e 
\nonumber \\
&& \qquad \qquad + 
      285000 \bth e + 1021440 c e - 3064320 d e - 1754352 e^2 + 
      331760 \aon \fon \nonumber \\
&& \qquad \qquad - 2970000 \atw \fon - 1047156 \bon \fon -
5824500 \btw \fon - 
      2953500 \bth \fon + 96000 c \fon \nonumber \\
&& \qquad \qquad - 288000 d \fon - 306160 e
\fon + 
      122436 \fon^2 - 2970000 \aon \ftw - 62986000 \atw \ftw \nonumber \\
&& \qquad \qquad -
2524500 \bon \ftw - 
      304434900 \btw \ftw - 6253500 \bth \ftw + 2400000 c \ftw \nonumber \\
&& \qquad \qquad -
7200000 d \ftw - 
      7654000 e \ftw + 285000 \fon \ftw + 6480900 \ftw^2) \nonumber \\
&& \qquad \qquad + {27\over 2} K^{-5} t_0 g^2 
\nonumber \\[3ex]
\VV_{134} &=& 
{1\over 16384} K^{-5} a_s \Big( \ron (196404 \aon - 283500 \atw  + 
298137 \bon  - 555975 \btw  - 
      121125 \bth  \nonumber \\
&& \qquad + 6720 c  - 20160 d  - 4788 e  + 21318 \fon
+ 
      57750 \ftw  + 132096 g  ) \nonumber \\
&& \qquad + \rtw (108500 \aon  - 1431500 \atw
+ 
      110625 \bon  - 6918975 \btw  + 767475 \bth  \nonumber \\
&& \qquad + 40000 c  - 
      120000 d  - 28500 e  - 74250 \fon  + 625350 \ftw -
384000 g   )\nonumber \\
&& \qquad + s 
      (-19096 \aon  
+ 297000 \atw  - 19470 \bon  + 582450 \btw  - 74250
\bth
\nonumber \\
&& \qquad - 
      9600 c  + 28800 d  + 30616 e  + 6156 \fon  - 28500 \ftw + 
      67584 g   ) \nonumber \\
&& \qquad 
+ y (305328 \aon  - 162000 \atw  + 502140 \bon  - 317700 \btw  + 
      40500 \bth  \nonumber \\
&& \qquad + 3840 c  - 11520 d  - 2736 e  - 7128 \fon  + 
      33000 \ftw  + 749568 g )  \Big)
\nonumber \\
\een

\sectiono{SFT Potential for Study of Lump Solution 
on a Circle} \label{a2} 

In this appendix we shall derive the string field theory potential
$\VVV(\Phi)$ relevant for studying the formation of the lump solution on a
circle of radius $R$. The expansion of the string field has been described
in section \ref{s3}. As mentioned there, since we are interested in
studying this phenomenon near $R=1$, in counting level of a field we shall
pretend as if $R$ has already been set to 1, although in the expression
for the potential we shall keep the complete 
$R$ dependence. We shall denote by
$\VVV^{(M,N)}$ the level $(M,N)$ approximation to $\VVV(\Phi)$, by
$\VVV_{mm}$ the quadratic term in the potential for level $m$ fields, and
by $\VVV_{mnp}$ the cubic term in the potential coupling a level $m$, a
level $n$ and a level $p$ field. As discussed 
in the text, using momentum
conservation it is easy to
show that with this definition of level, the total level of 
all the fields 
entering a term in the potential  must be even. We then have: 
\ben \label{esubeq} 
\VVV^{(0,0)} &=& \VVV_{00} + \VVV_{000}\nonumber \\
\VVV^{(1,2)} &=& \VVV^{(0,0)} + \VVV_{11} + \VVV_{011} \nonumber \\
\VVV^{(2,4)} &=& \VVV^{(1,2)} + \VVV_{22} 
+ \VVV_{002} + \VVV_{112} +\VVV_{022}
\nonumber \\
\VVV^{(3,6)} &=& \VVV^{(2,4)} + \VVV_{33} 
+ \VVV_{013} + \VVV_{222} + \VVV_{033}
+ \VVV_{123} \nonumber \\
\VVV^{(4,8)} &=& \VVV^{(3,6)} + \VVV_{44} + \VVV_{004} + \VVV_{114} +
\VVV_{024} + \VVV_{233} + \VVV_{224} 
+ \VVV_{044} + \VVV_{134} \nonumber\\
\een
Explicit computation gives the following expressions for $\VVV_{mm}$ and
$\VVV_{mnp}$:
\ben \label{emaineq}
\VVV_{00} &=& -{1\over 2} t_0^2 \nonumber \\
\VVV_{11} &=& - {1\over 4} (1 - R^{-2}) t_1^2 \nonumber \\
\VVV_{22} &=& -{1\over 2} u_0^2 + {1\over 4} v_0^2 + {25\over 4} w_0^2
\nonumber \\
\VVV_{33} &=& - {1\over 4} u_1^2 (1 + R^{-2}) + {1\over 8} v_1^2(1 +
R^{-2})
(1 + 8 R^{-2}) + {25\over 8} w_1^2 (1+ R^{-2}) \nonumber \\
&& + 3 R^{-2} v_1 z_1 (1 +
R^{-2}) +
R^{-2} z_1^2 (1 + R^{-2}) (1 + 2 R^{-2})\nonumber \\ \cr
\VVV_{44} &=& {15\over 2} \aon^2 
+ {375\over2} \atw^2 + 9 \aon \bon + {27\over
4}
\bon^2 + 225 \atw \btw + {2475\over 4} \btw^2 + {75\over 8} \bth^2 + 3 c d 
\nonumber \\
&& -
{3\over 2} e^2 - {3\over 4} \fon^2 -{75\over 4} \ftw^2 + {81\over 4} \wt g^2
-{1\over 4} (1 - 4 R^{-2}) t_2^2 \nonumber \\
\VVV_{000} &=& {1\over 3} K^3 t_0^3 \nonumber \\
\VVV_{011} &=& {1\over 2} K^{3 - 2 / R^{2}} t_0 t_1^2 \nonumber \\
\VVV_{002} &=& {1\over 32} K t_0^2\, (22 u_0 - 5 v_0 - 125 w_0) \nonumber \\
\VVV_{112} &=& K^{1 - 2/ R^{2}} t_1^2 \,\bigg( {11\over 32} u_0 +{1\over 2}
\Big(-{5\over 32} + R^{-2}\Big) v_0 -{125\over 64} w_0 \bigg)\nonumber \\
\VVV_{013} &=& {1\over 32} K^{1 - 2/R^2} t_0 t_1 \Big( 22 u_1 - (5 + 16
R^{-2}) v_1 - 125 w_1 + R^{-2} (-44 + 32 R^{-2}) z_1\Big)\nonumber \\
\VVV_{022} &=& {1\over 1024} K^{-1} t_0\, (228 u_0^2 - 220 u_0 v_0 + 537
v_0^2
- 5500 u_0 w_0 + 1250 v_0 w_0 + 28425 w_0^2) \nonumber \\
\VVV_{004} &=& {1\over 1024} K^{-1} t_0^2 \, (540 \aon 
+ 13500 \atw + 459 \bon +
26475 \btw +
625 \bth \nonumber \\
&& - 320 c + 960 d + 228 e - 110 \fon - 2750 \ftw)
\nonumber \\[3ex]
\VVV_{222}
& = & \displaystyle{
K \left\{ {1 \over 144} u_0^3 + {8321 \over 93312} 
v_0^3 - {219775 \over 10368} w_0^3  
- {95 \over 7776} u_0^2 
\left( v_0 + 25 w_0 \right) \right. } \nonumber \\
& & \displaystyle{
+ {1969 \over 15552} u_0 v_0^2 + 
{104225 \over 15552} u_0 w_0^2 - 
{22375 \over 31104} v_0^2 w_0} 
\displaystyle{\left. 
- {47375 \over 31104} v_0 w_0^2 
+ {6875 \over 23328} u_0 v_0 w_0 \right\} } \nonumber \\[3ex]
\VVV_{033} &=&  
\displaystyle{ K^{1 - 2/R^2} \left\{ {19 \over 288} t_0 u_1^2 
+ {1 \over 3456}  \left( 
537 + {8864 \over R^2} + {256 \over R^4} \right)  t_0 v_1^2 + {28425 
\over 3456}  t_0 w_1^2 \right.}
\nonumber \\
& & 
\displaystyle{
 - {11 \over 864} t_0 u_1 
\left( \left( 5 + {16 \over R^2} 
\right) v_1 + 125 w_1 \right) + {125 \over 1728}  \left( 5 + {16 \over
R^2} \right) 
t_0 v_1 w_1 } \nonumber \\
& & 
\displaystyle{
+ {11 \over 864} {1 \over R^2}\left( -44 + {32 \over R^2} \right)  
t_0 u_1 z_1 + {1 \over 432} \left( {2359 \over R^2} + 
{1672 \over R^4} - {128 \over R^6} \right) t_0 v_1 z_1} \nonumber \\
& & 
\displaystyle{ \left.
+ {125 \over 432} {1 \over R^2} \left( 11 - 
{8 \over R^2} \right) t_0 w_1 z_1
+ {1 \over 216}  { 1 \over R^2} 
\left( 384 + {1145 \over R^2} + 
{336 \over R^4} + {64 \over R^6} \right) t_0 z_1^2 \right\}}
\nonumber
\\[3ex]
\nonumber \\
\VVV_{123} &=& 
\displaystyle{ K^{1 - 2 / R^2} \bigg\{
{19 \over 144}  t_1 u_0 u_1 - {11 \over 864}  t_1 u_0 \left( \left( 5 +
{16 \over R^2}
\right) v_1 + 125 w_1 \right)} 
\nonumber \\
& & 
\displaystyle{
+ {1 \over 864} {1 \over R^2} \left( 11 \left( -44 + {32 \over R^2}
\right) 
t_1 u_0 z_1 + \left( 2158 - {2832 \over R^2} + {512 \over R^4} \right)  
t_1 v_0 z_1 \right)
} \nonumber \\
& &
\displaystyle{
- {11 \over 864} \left( 5 - {32 \over R^2} \right) 
t_1 v_0 u_1 + {1 \over 1728} \left( 
537 + {944 \over R^2} - {512 \over R^4} 
\right) t_1 v_0 v_1} \nonumber \\
& &
\displaystyle{ 
+ {25 \over 1728}  \left( 25 - {160 \over R^2} 
\right)  t_1 v_0 w_1 - {1375 \over 864} 
t_1 w_0 u_1 + {25 \over 1728} \left(
25 + {80 \over R^2} \right)  t_1 w_0 v_1} \nonumber \\ 
& & \displaystyle{
+ {28425 \over 1728}  t_1 w_0 w_1 
+ {125 \over 432} {1 \over R^2} \left( 11 - 
{8 \over R^2} \right)  t_1 w_0 z_1} \bigg\} \nonumber \\
\nonumber \\[3ex]
\VVV_{114} &=& K^{1 - 2 / R^2} t_1^2 \bigg\{ \Big( {5\over 32} - {11\over
27
R^2}\Big) \aon + {125 \over 32} \atw + \Big( {17\over 128} - {37\over 54
R^2}
+{8\over 27 R^4}\Big) \bon \nonumber \\
&& \qquad + {8825\over 1152} \btw + \Big( {625
\over 3456}
-{125\over 108 R^2} \Big) \bth - {5 \over 54} c + {5 \over 18} d + {19
\over
288} e \nonumber \\
&& \qquad  + \Big( -{55\over 1728} + {11\over 54 R^2} \Big) \fon 
- {1375 \over 1728} \ftw \bigg\} \nonumber \\
&& \qquad + {1\over 4} K^{3 - 6 / R^2} t_1^2 t_2 + {1\over 4}
K^{-1-2/R^2}R^{-2} 
(1 - 4 R^{-2}) t_1^2 \wt g
\nonumber \\[3ex]
\VVV_{024} &=& {1\over 16384} K^{-3} t_0 \Big( u_0
( 11880 \aon + 297000 \atw + 10098 \bon+ 582450 \btw +
13750 \bth - 
      9600 c \nonumber \\
&& \qquad \qquad + 28800 d + 30616 e - 1140 \fon - 28500 \ftw)
\nonumber \\
&& \qquad + v_0 ( 
      7540 \aon - 67500 \atw - 23799 \bon - 132375 \btw -
67125 \bth + 
      1600 c \nonumber \\ 
&& \qquad \qquad - 4800 d - 1140 e + 11814 \fon + 13750 \ftw)
\nonumber \\
&& \qquad + w_0
(- 
      67500 \aon - 1431500 \atw - 57375 \bon - 6918975 \btw - 
      142125 \bth \nonumber \\
&& \qquad \qquad + 40000 c - 120000 d - 28500 e + 13750 \fon + 
      625350 \ftw)\Big)
\nonumber \\
\VVV_{233} &=& {1 \over 65536} \, K^{-3 - 2 / R^2}  \Big(
1944 u_0 u_1^2 + (7296 R^{-2}- 1140) v_0 u_1^2 \nonumber \\
&& \qquad - 
    (7296 R^{-2} + 2280) u_0 u_1 v_1 +( - 22528 R^{-4} + 
    41536 R^{-2} + 23628) v_0 u_1 v_1 \nonumber \\
&& \qquad + (5632 R^{-4} + 
    195008 R^{-2} + 11814) u_0 v_1^2 \nonumber \\
&& \qquad + (8192 R^{-6} + 
    249600 R^{-4} - 233984 R^{-2} + 24963) v_0 v_1^2 - 
    28500 w_0 u_1^2 \nonumber \\
&& \qquad + (88000 R^{-2} + 27500) w_0 u_1 v_1 +( - 
    32000 R^{-4} - 1108000 R^{-2} - 67125) w_0 v_1^2 \nonumber \\
&& \qquad - 
    57000 u_0 u_1 w_1 + (- 176000 R^{-2} + 27500) v_0 u_1 w_1 + 
    (88000 R^{-2} + 27500) u_0 v_1 w_1 \nonumber \\
&& \qquad + (128000 R^{-4} - 
    236000 R^{-2} - 134250) v_0 v_1 w_1 + 1250700 w_0 u_1 w_1 \nonumber
\\
&& \qquad +( - 
    909600 R^{-2} - 284250) w_0 v_1 w_1 + 625350 u_0 w_1^2 + 
    (909600 R^{-2} - 142125) v_0 w_1^2 \nonumber \\
&& \qquad - 5933925 w_0 w_1^2 + 
    (14592 R^{-4} - 20064 R^{-2}) u_0 u_1 z_1 \nonumber \\
&& \qquad + (45056 R^{-6}
- 
    249216 R^{-4} + 189904 R^{-2}) v_0 u_1 z_1 \nonumber \\
&& \qquad +( - 22528
R^{-6} + 
    294272 R^{-4} + 415184 R^{-2}) u_0 v_1 z_1 \nonumber \\
&& \qquad +( - 32768
R^{-8} + 
    629760 R^{-6} - 790080 R^{-4} + 16232 R^{-2}) v_0 v_1 z_1 \nonumber
\\
&& \qquad +( - 
    176000 R^{-4} + 242000 R^{-2}) w_0 u_1 z_1 \nonumber \\
&& \qquad + (128000
R^{-6} - 
    1672000 R^{-4} - 2359000 R^{-2}) w_0 v_1 z_1 \nonumber \\
&& \qquad +( - 176000
R^{-4}
+ 
    242000 R^{-2}) u_0 w_1 z_1 \nonumber \\
&& \qquad +(- 256000 R^{-6} + 1416000
R^{-4}
- 
    1079000 R^{-2}) v_0 w_1 z_1 \nonumber \\
&& \qquad + (1819200 R^{-4} - 2501400
R^{-2}) w_0
w_1 z_1 \nonumber \\
&& \qquad 
+ 
    (22528 R^{-8} + 118272 R^{-6} + 403040 R^{-4} + 
    135168 R^{-2}) u_0 z_1^2 \nonumber \\
&& \qquad + (32768 R^{-10} - 95232 R^{-8} + 
    395520 R^{-6} - 517584 R^{-4} + 34816 R^{-2}) v_0 z_1^2 \nonumber \\
&& \qquad +
(- 
    128000 R^{-8} - 672000 R^{-6} - 2290000 R^{-4} - 
    768000 R^{-2}) w_0 z_1^2\Big)
\nonumber \\[3ex]
\VVV_{224} &=& {1\over 1048576} K^{-5} \Big(
u_0^2 (123120 \aon + 3078000 \atw + 104652 \bon 
+ 6036300 \btw + 
    142500 \bth \nonumber \\
&& \qquad - 103680 c + 311040 d +
1997584 e - 
    9720 \fon - 243000 \ftw) 
\nonumber \\
&& \qquad + u_0 v_0 ( 331760 \aon 
- 2970000 \atw - 
    1047156 \bon - 5824500 \btw - 2953500 \bth
\nonumber \\
&& \qquad
+ 96000 c - 
    288000 d - 306160 e + 244872 \fon 
+ 285000 \ftw) \nonumber \\
&& \qquad + v_0^2 (- 
    1254212 \aon + 7249500 \atw + 658131 \bon + 14217075 \btw - 
    3120375 \bth \nonumber \\
&& \qquad - 171840 c + 515520 d +
122436 e + 
    549186 \fon - 1476750 \ftw) \nonumber \\
&& \qquad + u_0 w_0 (- 2970000 \aon
- 
    62986000 \atw - 2524500 \bon - 304434900 \btw - 
    6253500 \bth \nonumber \\
&& \qquad + 2400000 c - 7200000 d - 7654000 e + 
    285000 \fon + 12961800 \ftw) \nonumber \\
&& \qquad + v_0 w_0 (- 1885000 \aon + 
    14315000 \atw + 5949750 \bon + 69189750 \btw + 
    30528450 \bth \nonumber \\
&& \qquad - 400000 c + 1200000 d + 285000 e - 
    2953500 \fon - 6253500 \ftw) \nonumber \\
&& \qquad +\, w_0^2\, (15349500 \aon + 
    283692700 \atw + 13047075 \bon + 1777831875 \btw + 
    29669625 \bth \nonumber \\
&& \qquad - 9096000 c + 27288000 d +
6480900 e - 
    3126750 \fon - 130546350 \ftw)\Big)
\nonumber \\[3ex]
\VVV_{044} &=&{1\over 1048576} K^{-5} t_0\,
(18846480 \aon^2 + 14580000 \aon \atw 
+ 646122000 \atw^2 + 21655656 \aon \bon 
\nonumber \\
&& \qquad \qquad + 
      12393000 \atw \bon + 10551033 \bon^2 + 28593000 \aon \btw 
+ 1120943400 \atw \btw \nonumber \\
&& \qquad \qquad + 
      24304050 \bon \btw + 2269335225 \btw^2 
- 1885000 \aon \bth + 14315000 \atw \bth \nonumber \\
&& \qquad \qquad + 
      5949750 \bon \bth + 69189750 \btw \bth 
+ 15264225 \bth^2 - 345600 \aon c - 
      8640000 \atw c \nonumber \\
&& \qquad \qquad - 293760 \bon c - 16944000 \btw c - 400000
\bth c + 
      76800 c^2 + 1036800 \aon d \nonumber \\
&& \qquad \qquad + 25920000 \atw d + 881280 \bon
d + 
      50832000 \btw d + 1200000 \bth d + 1636352 c d \nonumber \\
&& \qquad \qquad + 691200
d^2 + 
      246240 \aon e + 6156000 \atw e + 209304 \bon e + 12072600 \btw e 
\nonumber \\
&& \qquad \qquad + 
      285000 \bth e + 1021440 c e - 3064320 d e - 1754352 e^2 + 
      331760 \aon \fon \nonumber \\
&& \qquad \qquad - 2970000 \atw \fon - 1047156 \bon \fon -
5824500 \btw \fon - 
      2953500 \bth \fon + 96000 c \fon \nonumber \\
&& \qquad \qquad - 288000 d \fon - 306160 e
\fon + 
      122436 \fon^2 - 2970000 \aon \ftw - 62986000 \atw \ftw \nonumber \\
&& \qquad \qquad -
2524500 \bon \ftw - 
      304434900 \btw \ftw - 6253500 \bth \ftw + 2400000 c \ftw \nonumber \\
&& \qquad \qquad -
7200000 d \ftw - 
      7654000 e \ftw + 285000 \fon \ftw + 6480900 \ftw^2) \nonumber \\
&& \qquad \qquad + {27\over 2} K^{-5} t_0 \wt g^2 + {1\over 2} K^{3 -
8/R^2} t_0 t_2^2 
\nonumber \\[3ex]
\VVV_{134} &=& {1\over 32768} K^{-3 - 2 / R^2} t_1 
\Big( (-30976 R^{-2}
+ 11880) \aon u_1 + 297000 \atw u_1 \nonumber \\
&& \qquad + 
      (22528 R^{-4} - 52096 R^{-2} + 10098) \bon u_1 + 
      582450 \btw u_1 \nonumber \\
&& \qquad +( - 88000 R^{-2} + 13750) \bth u_1 - 
      9600 c u_1 + 28800 d u_1 + 30616 e u_1 \nonumber \\
&& \qquad + (7296 R^{-2} - 
      1140) \fon u_1 - 28500 \ftw u_1 \nonumber \\
&& \qquad + (22528 R^{-4} + 
      166336 R^{-2} + 7540) \aon v_1 +( - 216000 R^{-2} - 
      67500) \atw v_1 \nonumber \\
&& \qquad +( - 16384 R^{-6} + 98304 R^{-4} + 
      240016 R^{-2} - 23799) \bon v_1 \nonumber \\
&& \qquad +(- 423600 R^{-2} - 
      132375) \btw v_1 + (64000 R^{-4} - 118000 R^{-2} - 
      67125) \bth v_1 \nonumber \\
&& \qquad + (5120 R^{-2} + 1600) c v_1 +(- 15360
R^{-2} - 
      4800) d v_1 +(- 3648 R^{-2} - 1140) e v_1 \nonumber \\
&& \qquad +( - 11264
R^{-4} + 
      20768 R^{-2} + 11814) \fon v_1 + (44000 R^{-2} + 
      13750) \ftw v_1 \nonumber \\
&& \qquad + (176000 R^{-2} - 67500) \aon w_1 - 
      1431500 \atw w_1 \nonumber \\
&& \qquad +(- 128000 R^{-4} + 296000 R^{-2} - 
      57375) \bon w_1 - 6918975 \btw w_1 \nonumber \\
&& \qquad + (909600 R^{-2} - 
      142125) \bth w_1 + 40000 c w_1 - 120000 d w_1 - 
      28500 e w_1 \nonumber \\
&& \qquad +(- 88000 R^{-2} + 13750) \fon w_1 + 
      625350 \ftw w_1 \nonumber \\
&& \qquad +(- 45056 R^{-6} + 193920 R^{-4} + 
      156464 R^{-2}) \aon z_1 \nonumber \\
&& \qquad + (432000 R^{-4} - 594000 R^{-2})
\atw
z_1 \nonumber \\
&& \qquad + 
      (32768 R^{-8} - 382976 R^{-6} + 880736 R^{-4} - 
      28388 R^{-2}) \bon z_1 \nonumber \\
&& \qquad + (847200 R^{-4} - 1164900 R^{-2})
\btw
z_1 \nonumber \\
&& \qquad +( - 
      128000 R^{-6} + 708000 R^{-4} - 539500 R^{-2}) \bth z_1 \nonumber \\
&& \qquad - 
      (10240 R^{-4} - 14080 R^{-2}) c z_1 + (30720 R^{-4} - 
      42240 R^{-2}) d z_1 \nonumber \\
&& \qquad + (7296 R^{-4} - 10032 R^{-2}) e z_1
+ 
      (22528 R^{-6} - 124608 R^{-4} + 94952 R^{-2}) \fon z_1 \nonumber \\
&& \qquad +(
- 
      88000 R^{-4} + 121000 R^{-2}) \ftw z_1\Big) \nonumber \\
&& + {1\over 64} K^{-3 - 2 / R^2} t_1 \wt g R^{-4}
(-2 + R) (2 + R) \Big(22 u_1 + (- 16 R^{-2} + 59) v_1 - 125 w_1 \nonumber 
\\
&& \qquad + 
      (32 R^{-4} - 300 R^{-2} + 512) z_1\Big) \nonumber \\
&& + {1 \over 64} K^{1 - 6/R^2}
t_1 t_2 \Big( 22 u_1 + (48 R^{-2} -
5) v_1 - 125 w_1 + 
      (288 R^{-4} - 44 R^{-2}) z_1\Big)
\nonumber \\
\een

\end{document}